\documentclass[pdflatex,sn-mathphys]{sn-jnl}

\jyear{2022}%

\theoremstyle{thmstyleone}%
\theoremstyle{thmstyletwo}%

\theoremstyle{thmstylethree}%

\usepackage{graphicx}
\begin{document}

\title[Trajectories of spinning test particles]{The Ricci Rotation Coefficients in the description of trajectories of spinning test particles off-equatorial planes in a rotational gravitational field}


\author*[1]{\fnm{Nelson} \sur{Velandia SJ}}\email{navelandia@javeriana.edu.co}

\author[1]{\fnm{Alfonso} \sur{Leyva}}\email{leyvaa@javeriana.edu.co}

\author[2]{\fnm{Javier Alexander} \sur{Cano-Arango}}\email{javier.cano@correo.nucleares.unam.mx}

\affil*[1]{\orgdiv{Departmento de Fisica}, \orgname{Facultad de Ciencias, Pontificia Universidad Javeriana}, \orgaddress{\street{Cr. 7 No 40 - 62} \city{Bogotá D.C.}, \postcode{110231}, \state{} \country{Colombia}}}

\affil[2]{\orgdiv{Departamento de Física de Altas Energías, Instituto de Ciencias Nucleares}, \orgname{Universidad Nacional Autónoma de México}, \orgaddress{\street{Apartado Postal 70-543} \city{Ciudad de México}, \postcode{110231}, \state{} \country{Colombia}}}


\abstract{We describe the trajectories of circular orbits of spinless and spinning test particles around of rotating bodies in equatorial and non-equatorial planes via the Mathisson-Papapetrou-Dixon equations which include the Ricci rotation coefficients with the purpose of describing not only the curvature of space time, but also the rotation of the spinning test particles that orbit around the rotating masive bodies.}




\maketitle

\section{Introduction}\label{sec1}

    For to study the physics of a rotational gravitational field, one of the ways is describe the motion of test particles in this field. Tipically in the literature, the description of the motion of spinning test particles in a rotating gravitational field in non-equatorial plane we finds two main approaches. The first one is given by Mathisson-Papapetrou-Dixon Equations (MPD)  \cite{Papapetrou}, \cite{dixon}, \cite{mathisson} which yields the equations of motion of a spinning test particle in a given gravitational field. The second takes the first integrals of motion given by Carter \cite{carter} and derives the orbits of spinless test particles around of rotating massive bodies. The term of particle is because the size of the test body is small compared to the scale of the curvature.
    The motion of particles on the metric type Kerr have been discussed by many authors \cite{bardeen}, \cite{dadhich}, \cite{bini-2005}, \cite{tartaglia}. Many of these works study the motion in the equatorial plane for spinless and spinning test particles via MPD Equations \cite{tanaka-1996}, \cite{suzuki}. Using this last formalism, we calculate the trajectories of spinless and spinning test particles in non-equatorial planes. On the other hand, in the literature, there are studies that use the Carter´s equations for both spinless and spinning test particles in equatorial planes \cite{abramowicz}, \cite{tod}, but there are few works that describe the motion of spinless test particles around on rotating bodies in non-equatorial planes \cite{wilkins}, \cite{tsoubelis}. This is the first novelty in relation to previous works. On the other hand, the study of trajectories of spinning test particles in non-equatorial planes for a Kerr spacetime under the Carter´s equations needs further investigation.
    We focus our work in the formalism given by the MPD equations \cite{dixon}. In this via, it is given a distribution of mass ($m$) with spin tensor $ (S^{\rho \sigma })$ around on central source ($M$) which has a metric tensor $ g_{\mu \nu}$. These equations of motion for a spinning test particle are obtained in terms of an expansion that depends on the derivatives of the metric and the multipole moments of the energy-momentum tensor $(T^{\mu \nu})$ \cite{dixon}, and are given by

\begin{eqnarray}
\frac{D}{d\tau }p^{\mu }\left( \tau \right) &=&-\frac{1}{2}R^{\mu }\text{ }
_{\nu \rho \sigma }v^{\nu }\left( \tau \right) S^{\rho \sigma }\left( \tau
\right)  \label{mov1} \\
\frac{D}{d\tau }S^{\mu \nu }\left( \tau \right) &=&2p^{[\mu }\left( \tau
\right) v^{\nu ]}\left( \tau \right) ,  \label{mov}
\end{eqnarray}
where $D/d\tau $ means the covariant derivative, the vector $p^{\mu }$ and
the antisymmetric tensor $S^{\mu \beta }$ are the linear and spin angular
momenta respectively. $R^{\mu }{}_{\nu \rho \sigma }$ is the curvature
tensor. Since $p_{\mu }p^{\mu }=$ constant and $S_{\rho \sigma }S^{\rho
\sigma }=$ constant along the particle trajectory \cite{wald}; then, we may
set

\begin{equation}
p^{\mu }=mu^{\mu }\text{, \ \ \ \ \ \ \ \ }u_{\mu }u^{\mu }=-1
\end{equation}

\begin{equation}
S^{\rho \sigma }=\epsilon ^{\rho \sigma }\text{ }_{\mu \nu }p^{\mu }S^{\nu }
\end{equation}

\begin{equation}
S^{2}=S_{\mu }S^{\mu }=\frac{1}{2m^{2}}S_{\mu \nu }S^{\mu \nu },
\end{equation}
and with these expressions, we obtain the center of mass condition and the
relation between the spin tensor and the vector spin which is called spin
supplementary condition (SSC). For this case, we used the Tulczyjew
supplementary condition \cite{kyrian}:
\begin{equation}
p_{\mu }S^{\mu }=0.
\end{equation}

In the present paper, we calculate numerically, via MPD equations, the
trajectories for spinless and spinning test particles in circular orbits
when the test particles are in non-equatorial planes. We are interested in
to study the effects from spin of the test particle orbiting in a rotating
gravitational field which is generated by a massive rotating body. In the
majority of cases, the description of trajectories is given by the study of
the Christoffel symbols, the novelty of our work is to take the Ricci
Rotation Coefficients (RRC) in orden to study the trajectories of spinning
test particles in non-equatorial planes. On the other hand, when we study
these trajectories, we can describe the space time around a rotating massive
body and its phenomena.

The Carter\'{}s equations are given by the first integrals of the equations of motion of a
spinless test particle around on a rotating body for a Kerr metric both in
an equatorial plane and in non-equatorial planes \cite{carter}. These
equations use the symmetries of the Kerr geometry and the conserved
quantities of energy ($E$), angular momentum ($L$), rest mass ($M$) and a
fourth integral of motion called Carter\'{}s constant ($Q$). For the case of spinning test particles in the equatorial
plane, when the space-time possess a Killing field, there exists a linear
combination of the components of momenta ($p^{\mu }$) and the spin tensor ($
S^{\mu \nu }$) \cite{dixon-1979}. The study for orbits of spinning test
particles off to equatorial plane by the Carter\'{}s equations needs further investigation.

We shall compare the results by numerical integration in three cases: First,
we compare between the trajectories of spinless test particles in non
equatorial planes given by the Carter\'{}s equations and the MPD equations. Second, the trajectories both of spinless
and spinning test particles in non-equatorial planes given by the MPD
equations. Third, we compare the behavior of two Boyer Linquist coordinates (
$\theta $, $\varphi $) in regard to the proper time ($\tau $) both to
spinless and to spinning test particles.

The paper is organized as follows. In Section 2, the equations of motion MPD
are reduced to expressions that include the Ricci Rotation Coefficients
(RRC). These equations describe the motion of spinless and spinning test
particles around on a rotating body. We study the cases when the spinless
and spinning test particles are in the equatorial plane and give the set of
equations when the spinless and spinning test particles are out of the
equatorial plane. In Section 3, there is a brief description of the Carter\'{}s equations for spherical orbits. These equations take three constants of
motion given by the symmetries of a rotating body. A fourth constant is
obtained from the separability of the Hamilton-Jacobi equation. These orbits
are calculated numerically. Then, in Section 4, we present a comparison
numerical between the MPD equations and Carter\'{}s equations for the spinless test particles both in the equatorial plane and
the non-equatorial plane. We take the initial values from the Carter\'{}s equations and are replaced in the MPD equations. In the last section, the
conclusions and some futures works are formulated in order to describe
spinning test particles in a Kerr type metric.

\section{MPD equations and the Ricci Rotation
Coefficients}\label{sec2}
The equations MPD had been traditionally used in the equatorial plane. We
focus our study not only in the equatorial plane, but also in the non
equatorial plane. In this paper, we take the work from Tanaka \textit{et al.}
\cite{tanaka-1996} who reduce the equations of motion using the RRC. They
solve the equations of motion for a circular orbit in the equatorial plane
while we, with the same via, solve the case for trajectories non-equatorial
plane. Given this, the equations of motion (\ref{mov1}) and (\ref{mov}) for
the tetrad components\ are reduced to

\begin{eqnarray}
\frac{du^{\alpha }}{d\tau } &=&\omega _{\beta \gamma }{}^{\alpha }v^{\beta
}u^{\gamma }-SR^{\alpha },  \label{4.8a} \\
\frac{d\zeta ^{\alpha }}{d\tau } &=&\omega _{\beta \gamma }{}^{\alpha
}v^{\beta }\zeta ^{\gamma }-Su^{\alpha }\zeta ^{\beta }R_{\beta }.
\label{4.8c}
\end{eqnarray}

\begin{equation}
v^{\mu }-u^{\mu }=\frac{1}{2}\left( \frac{S^{\mu \nu }R_{\nu \rho \sigma
\kappa }u^{\rho }S^{\sigma \kappa }}{m^{2}+\frac{1}{4}R_{\chi \xi \zeta \eta
}S^{\chi \xi }S^{\zeta \eta }}\right) ,  \label{4.8d}
\end{equation}
where $\omega _{\beta \gamma }{}^{\alpha }$ are the RRC and $\zeta ^{\alpha
} $ is the unit spin vector, which is defined by $\zeta ^{\alpha }:=S^{a}/S$
. Here $R^{\alpha }$ and $S^{\mu \nu }$ are defined by

\begin{equation}
R^{\alpha }:=R^{\ast \alpha }{}_{\beta \gamma \lambda }v^{\beta }u^{\gamma
}\zeta ^{\lambda }=\frac{1}{2mS}R^{\alpha }{}_{\beta \gamma \lambda
}v^{\beta }S^{\gamma \lambda }.  \label{4.8b}
\end{equation}

\begin{equation}
S^{\gamma \lambda }=mS\epsilon ^{\gamma \lambda }\text{ }_{\vartheta \kappa
}u^{\vartheta }\zeta ^{\kappa }=m\epsilon ^{\gamma \lambda }\text{ }
_{\vartheta \kappa }u^{\vartheta }S^{\kappa },
\end{equation}
the last equation yields the relation between tensor spin ($S^{\gamma
\lambda }$) and the vector spin ($S^{\kappa }$). For our calculation, we
have an orthonormal frame with the RRC and we can extract information about
the curvature of the spacetime in question. It will be important when we
describe the trajectories of test particles off-equatorial plane.

On the other hand, the RRC are related with the the coordinates of the
derivative vectors that measure not only the deviation, but also the
rotation of all the frame vector when moved in various directions.
Furthermore, we can observe that the RRC carry information about how the
frame rotates and not just information about the curvature of space time.

\subsection{Calculation of the Ricci Rotation Coefficients}

Next, it is to define the Ricci rotation coefficients for a Kerr spacetime 
\cite{mino} as $\omega _{ab}{}^{c}=e_{a}{}^{\mu }e_{b}{}^{\nu }e^{c}{}_{\nu
;\mu }$ where the semicolon indicates the covariant derivative (Appendix A).

In order to find a solution, it is convenient to introduce the tetrad frame 
\cite{tanaka-1996} defined by $e^{i}{}_{\mu }=\left(
e^{i}{}_{t},e^{i}{}_{r},e^{i}{}_{\theta },e^{i}{}_{\varphi }\right) $

\begin{eqnarray}
e^{0}{}_{\mu } &=&\left( \sqrt{\frac{r^{2}-\frac{2GMr}{c^{2}}+a^{2}}{
r^{2}+a^{2}\cos ^{2}\theta }},0,0,-a\sin ^{2}\theta \sqrt{\frac{r^{2}-\frac{
2GMr}{c^{2}}+a^{2}}{r^{2}+a^{2}\cos ^{2}\theta }}\right) ,  \notag \\
e^{1}{}_{\mu } &=&\left( 0,\sqrt{\frac{r^{2}+a^{2}\cos ^{2}\theta }{r^{2}-
\frac{2GMr}{c^{2}}+a^{2}}},0,0\right) ,  \notag \\
e^{2}{}_{\mu } &=&\left( 0,0,\sqrt{r^{2}+a^{2}\cos ^{2}\theta },0\right) , 
\notag \\
e^{3}{}_{\mu } &=&\left( -\frac{a\sin \theta }{\sqrt{r^{2}+a^{2}\cos
^{2}\theta }},0,0,\frac{\left( r^{2}+a^{2}\right) \sin \theta }{\sqrt{
r^{2}+a^{2}\cos ^{2}\theta }}\right)  \label{4.1}
\end{eqnarray}
where the Greek letters distinguish the tensor indices ($\mu =t,r,\theta
,\varphi $) from the tetrad indices which are in the Latin alphabet ($
i=0,1,2,3$). Here it is used the metric signature $\left( -,+,+,+\right) $.

The RRC are given by \cite{laranaga}, \cite{chandra}
\begin{equation}
\omega _{(a)(b)(c)}=\frac{1}{2}\left[ \lambda _{(a)(b)(c)}+\lambda
_{(c)(a)(b)}-\lambda _{(b)(c)(a)}\right]  \label{wabc}
\end{equation}
where $\lambda _{(a)(b)(c)}$ are the rotation pre-coefficients and are
defined by
\begin{equation}
\lambda _{(a)(b)(c)}=e_{(b)\mu ,\nu }\left[ e_{\left( a\right) }{}^{\mu
}e_{\left( c\right) }{}^{\nu }-e_{\left( a\right) }{}^{\nu }e_{\left(
c\right) }{}^{\mu }\right]  \label{labc}
\end{equation}

\subsection{Spinless test particles in a Kerr metric}

In this section, the equations of motion for spinless test particles in a
Kerr metric with radius constant ($r=r_{0}$) are solved. The aim is to write
down the equations with the help of the Eq. (\ref{4.8a}) and the RRC
(Appendix A). In this case, the magnitude of the spin is equal to zero ($S=0$
) and $r$ is constant, i.e. $u^{1}=0$.\ Since the test particles is spinless
and according to Eq. (\ref{4.8d}), it can be identified the velocity $v^{\mu
}$ with the velocity $u^{\mu }$ in Eq. (\ref{4.8a}). Therefore the system of
equations is given by

\begin{eqnarray}
\frac{d}{d\tau }u^{0} &=&\omega _{02}{}^{0}u^{0}u^{2}  \label{4.11a} \\
\frac{d}{d\tau }u^{1} &=&\omega _{02}{}^{1}u^{0}u^{2}+2\omega
_{03}{}^{1}u^{0}u^{3}+\omega _{22}{}^{1}u^{2}u^{2}+\omega
_{33}{}^{1}u^{3}u^{3}=0  \label{4.11b} \\
\frac{d}{d\tau }u^{2} &=&\omega _{00}{}^{2}u^{0}u^{0}+2\omega
_{03}{}^{2}u^{0}u^{3}+\omega _{33}{}^{2}u^{3}u^{3}  \label{4.11c} \\
\frac{d}{d\tau }u^{3} &=&-2\omega _{02}{}^{3}u^{0}u^{2}-\omega
_{23}{}^{3}u^{2}u^{3}.  \label{4.11d}
\end{eqnarray}

Let be replaced the values of the RRC and the tetrads in Eqs. (\ref{4.11a})
- (\ref{4.11d}). These latter equations for spinless test particles will be
solved by numerical integration with Mathematica.

For orbits in the equatorial plane, we assume that corresponds to: $r=$
constant ($u^{1}=0$), and close to the equatorial plane ($u^{2}=O(\overset{
\sim }{\theta )}$), that is, $\overset{\sim }{\theta }:=\theta -\pi
/2=O\left( S/M\right) \ll 1$. Under this assumption, the nontrivial equation
is \cite{tanaka-1996}
\begin{equation}
\frac{d}{d\tau }u^{1}=\omega _{00}{}^{1}u^{0}u^{0}+\omega _{30}\text{ }%
^{1}u^{3}u^{0}+\omega _{03}{}^{1}u^{0}u^{3}+\omega _{33}{}^{1}u^{3}u^{3}=0
\end{equation}

In this case, the orbital angular velocity is given by
\begin{equation}
\Omega =\frac{\pm \sqrt{M}}{r^{{{}^3}/{{}^2}
}\pm \sqrt{M}a}
\end{equation}
where $M$ is the mass of central body and $a$ is the angular momentum of the
source.

\subsection{Spinning test particles in a Kerr metric}

First of all, for the case of spinning test particles in circular orbits and
in the equatorial plane, we assume that corresponds to: $r=$ constant ($
u^{1}=0$), close to the equatorial plane ($u^{2}=O$($\overset{\sim }{\theta 
\text{)}}$), and with the magnitude of spin ($S^{2}=S_{\mu }S^{\mu }$). In
this case, according to Eq. (\ref{4.8a}), the nontrivial equations of the
orbital motion are \cite{tanaka-1996}
\begin{eqnarray}
\frac{d}{d\tau }u^{1} &=&\omega _{00}{}^{1}u^{0}u^{0}+\omega _{30}\text{ }
^{1}u^{3}u^{0}+\omega _{03}{}^{1}u^{0}u^{3}+\omega
_{33}{}^{1}u^{3}u^{3}-SR^{1}=0  \notag \\
\frac{d}{d\tau }u^{2} &=&\left( \omega _{00}\text{ }^{2}u^{0}u^{0}+\omega
_{30}\text{ }^{2}u^{3}u^{0}+\omega _{03}\text{ }^{2}u^{0}u^{3}+\omega _{33}
\text{ }^{2}u^{3}u^{3}\right) \overset{\sim }{\theta }-SR^{2}
\end{eqnarray}%
where the components of $R^{a}$ are given by
\begin{equation*}
R^{0}=R^{3}=O(\overset{\sim }{\theta )}
\end{equation*}

\begin{eqnarray}
R^{1} &=&3\frac{M}{r^{3}}u^{0}u^{3}\frac{S^{2}}{S}+O(\overset{\sim }{\theta )
}  \notag \\
R^{2} &=&3\frac{M}{r^{3}}u^{0}u^{3}\frac{S^{1}}{S}+O(\overset{\sim }{\theta )
}
\end{eqnarray}

In the equatorial plane, the orbital angular velocity for spinning test
particles is given by
\begin{equation}
\Omega =\pm \frac{\sqrt{M}}{r^{{{}^3}/{{}^2}
}\pm a\sqrt{M}}\left[ 1-\frac{3S}{2}\frac{\pm \sqrt{Mr}-a}{r^{2}\pm a\sqrt{Mr
}}\right] +O(\overset{\sim }{\theta )}
\end{equation}
this new expression corresponds to the spin contribution ($S$) at $O(\overset
{\sim }{\theta )}$.

Now for the orbits non-equatorial, first, we take the work by Tanaka \textit{
et al. }\cite{tanaka-1996} which studies the motion of the spinning test
particle in the equatorial plane, and then, we go one step further and
numerically calculate the trajectories off-equatorial orbits. For this
treatment, we replace the "dynamical velocity $v$" Eq (\ref{4.8d}) in Eqs. (
\ref{4.8a}) and (\ref{4.8c}), and obtain the set of equations:

\begin{eqnarray}
\frac{du^{\alpha }}{d\tau } &=&\omega _{\beta \gamma }\text{ }^{\alpha
}\left( u^{\beta }+\frac{\epsilon ^{\beta \nu }\text{ }_{\varphi \lambda
}u^{\varphi }S^{\lambda }R_{\nu \rho \sigma \kappa }u^{\rho }\epsilon
^{\sigma \kappa }\text{ }_{\tau \upsilon }u^{\tau }S^{\upsilon }}{8+2R_{\chi
\xi \zeta \eta }\epsilon ^{\chi \xi }\text{ }_{\iota \varphi }u^{\iota
}S^{\varphi }\epsilon ^{\zeta \eta }\text{ }_{\delta \varepsilon }u^{\delta
}S^{\varepsilon }}\right) u^{\gamma }  \notag \\
&&-\frac{1}{2}g^{\alpha \eta }R_{\eta \mu \rho \pi }\left( u^{\mu }+\frac{
\epsilon ^{\mu \nu }\text{ }_{\gamma \lambda }u^{\gamma }S^{\lambda }R_{\nu
\rho \sigma \kappa }u^{\rho }\epsilon ^{\sigma \kappa }\text{ }_{\omega
\vartheta }u^{\omega }S^{\vartheta }}{8+2R_{\chi \xi \zeta \eta }\epsilon
^{\chi \xi }\text{ }_{\iota \varphi }u^{\iota }S^{\varphi }\epsilon ^{\zeta
\eta }\text{ }_{\delta \varepsilon }u^{\delta }S^{\varepsilon }}\right)
\epsilon ^{\rho \pi }\text{ }_{\sigma \kappa }u^{\sigma }S^{\kappa }  \notag
\\
&&  \label{s1}
\end{eqnarray}

\begin{eqnarray}
\frac{dS^{\alpha }}{d\tau } &=&\omega _{\beta \gamma }\text{ }^{\alpha
}\left( u^{\beta }+\frac{\epsilon ^{\beta \nu }\text{ }_{\varphi \lambda
}u^{\varphi }S^{\lambda }R_{\nu \rho \sigma \kappa }u^{\rho }\epsilon
^{\sigma \kappa }\text{ }_{\tau \upsilon }u^{\tau }S^{\upsilon }}{8+2R_{\chi
\xi \zeta \eta }\epsilon ^{\chi \xi }\text{ }_{\iota \varphi }u^{\iota
}S^{\varphi }\epsilon ^{\zeta \eta }\text{ }_{\delta \varepsilon }u^{\delta
}S^{\varepsilon }}\right) S^{\gamma }  \notag \\
&&-\frac{1}{2}u^{\alpha }S^{\beta }R_{\beta \rho \lambda \mu }\left( u^{\rho
}+\frac{\epsilon ^{\rho \nu }\text{ }_{\varphi \lambda }u^{\varphi
}S^{\lambda }R_{\nu \rho \sigma \kappa }u^{\rho }\epsilon ^{\sigma \kappa }
\text{ }_{\tau \upsilon }u^{\tau }S^{\upsilon }}{8+2R_{\chi \xi \zeta \eta
}\epsilon ^{\chi \xi }\text{ }_{\iota \varphi }u^{\iota }S^{\varphi
}\epsilon ^{\zeta \eta }\text{ }_{\delta \varepsilon }u^{\delta
}S^{\varepsilon }}\right) \epsilon ^{\lambda \mu }\text{ }_{\phi \delta
}u^{\phi }S^{\delta }  \notag \\
&&  \label{s2}
\end{eqnarray}

For the numerical integration, we need the initial values of the four
velocity which will be obtained from the Carter\'{}s equations.

\section{Equations of Motion and Carter\'{}s constant}

In a manifold type Kerr, the symmetries provide three constants of motion:
the energy ($E$), the angular momentum ($L$), and the rest central mass ($M$
). In addition, there is another, one constant which is due by the
separability of the Hamilton - Jacobi equation and is called Carter%
\'{}s constant ($Q$). The equation of Lagrange for a Kerr metric leads
immediately to the first integrals of the $t$ and $\varphi $ equations and
for the other two integrals of ($r$) and ($\theta $) are obtained by a
separable solution of the Hamilton - Jacobi equation. The set of equations
is given by \cite{stog}

\begin{eqnarray}
\Sigma \overset{\cdot }{t} &=&a\left( L-aE\sin ^{2}\theta \right) +\frac{
r^{2}+a^{2}}{\Delta }\left[ E\left( r^{2}+a^{2}-aL\right) \right] ,
\label{1} \\
\Sigma \overset{\cdot }{r}^{2} &=&\left[ E\left( r^{2}+a^{2}\right) \mp aL
\right] ^{2}-\Delta \left[ r^{2}+Q+\left( L\mp aE\right) ^{2}\right] ,
\label{2} \\
\Sigma \overset{\cdot }{\theta }^{2} &=&Q-\cos ^{2}\theta \left[ a^{2}\left(
1-E^{2}\right) +\frac{L^{2}}{\sin ^{2}\theta }\right] ,  \label{3} \\
\overset{\cdot }{\Sigma \varphi }\text{ } &=&\frac{L}{\sin ^{2}\theta }-aE+
\frac{a}{\Delta }\left[ E\left( r^{2}+a^{2}\right) -aL\right] ,  \label{4}
\end{eqnarray}
where $L$, $E$, and $Q$ are constants and

\begin{eqnarray*}
\Sigma &:&=r^{2}+a^{2}\cos ^{2}\theta , \\
\Delta &:&=r^{2}+a^{2}-2Mr,
\end{eqnarray*}%
$M$, and $a=J/Mc$ denote the central mass and specific angular momentum of
the central source which gives rise to the gravitational field represented
by the Kerr space-time.

In the study of orbits of test particles around on Kerr metric, there exists
a kind of orbits called spherical, \textit{i.e.}, constant radius. This last
orbit intersects the equatorial plane in a point called node. Since the
metric has angular momentum, the nodes of spherical orbits are dragged in
the sense of the spin of the rotating body. When there is a particle
orbiting in a nonequatorial orbit, this traces a kind of helix until a
maximum of latitude and when reachs the maximum this begins to descend until
a minimum latitude which is symmetric to the maximum \cite{kheng}.

When the space-time admits a Killing vector $\xi ^{\upsilon }$, there exists
a property that includes the covariant derivative and the spin tensor, which
gives a constant and is given by the expression \cite{dixon-1979}
\begin{equation}
p^{\nu }\xi _{\nu }+\frac{1}{2}\xi _{\nu ,\mu }S^{\nu \mu }=\text{ constant,}
\label{16}
\end{equation}
where $p^{\nu }$ is the linear momentum, $\xi _{\nu ,\mu }$ is the covariant
derivative of Killing vector, and $S^{\nu \mu }$ is the spin tensor of the
particle. In the case of Kerr metric, there are two Killing vectors, owing
to its stationary and axisymmetric nature. As consequence, Eq. (\ref{16})
yields two constants of motion: $E$ is the total energy and $L$ is the
component of its angular momentum along the axis of symmetry. Unfortunately,
there is not a constant of motion associated to the coordinate $r$;
therefore, for this way, there does not exist a equation of motion that
connects the constant of motion linked to coordinate $r$ and the spin
tensor. This mathematical expression needs further investigation.

On the other hand, people avoid the error that shows the numerical
integration due to the explicit square roots in the $r$ and $\theta $ in
Eqs.(\ref{2}) and (\ref{3}). In this formulation, they take the Hamilton%
\'{}s canonical equations which denote differentiation of the Hamiltonian with
respect to $r$ and $\theta $. This demostration is based on a taxonomy of
all periodic orbits around black holes \cite{levin}. The set of equations
constitutes a smooth system of ordinary differential equations and can be
integrated without change of variables. These equations were integrated
numerically to generate the periodic orbits. These orbits describe the
motion of test particles out of equatorial plane, but this approach does not
take into account the spinning test particles. So far, there is not a work
that connect successfully the equations of motion given by Carter and spin
of the test particle. Within Carter\'{}s equations, one finds in the literature a number of works, but none
considers the spinning test particle out of the equatorial plane \cite{iorio}.

\section{Numerical integration of the Equations of motion}

\begin{figure}[h]
\centering
\includegraphics[width=0.6\textwidth]{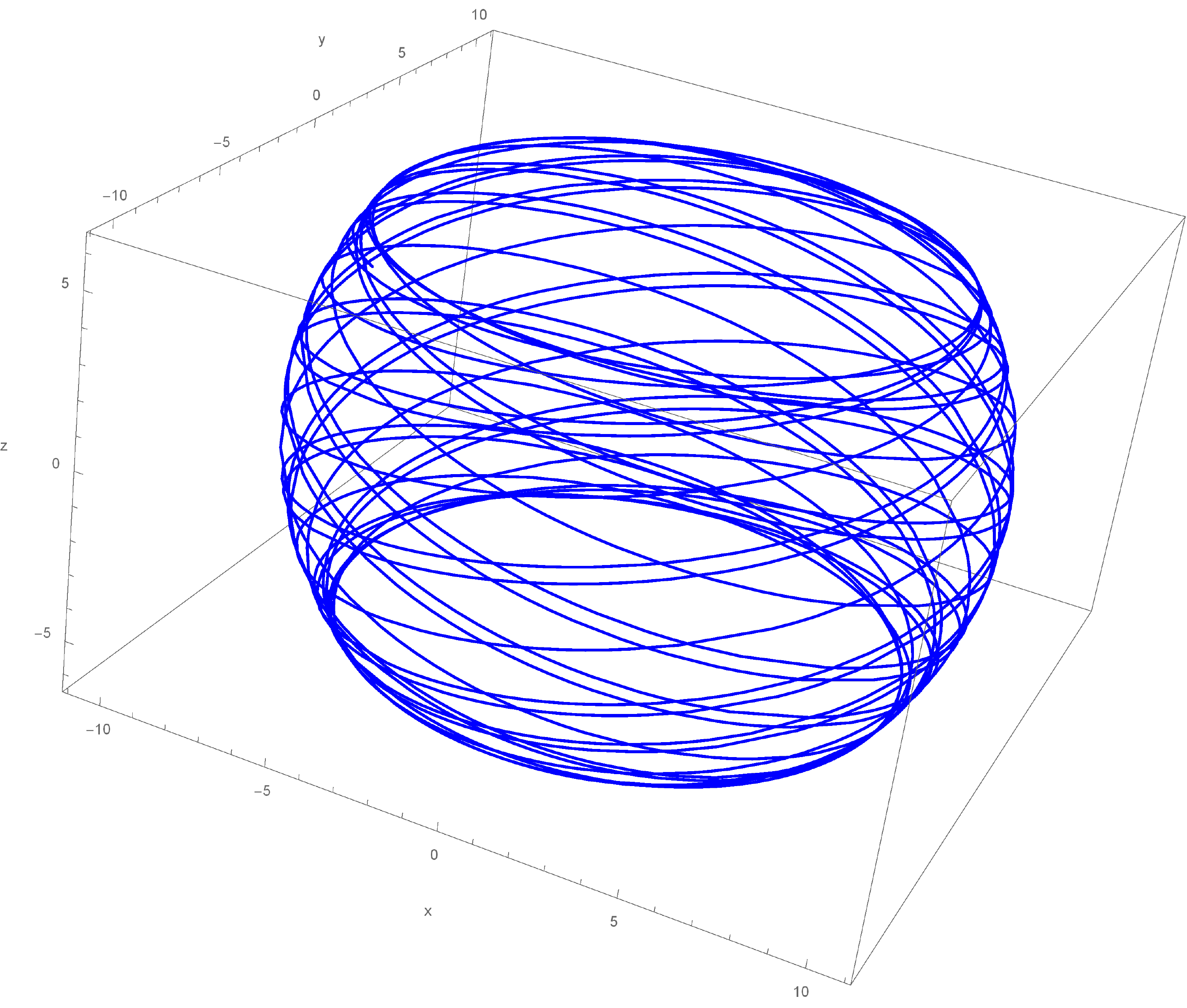}
\caption{Trajectory for a spinless test
particle (blue color). The values of the parameters are: $r=10$, $M=1$, $
E=0.9525$, $a=1$, $Q=4.224806$, $L=2.810974$}\label{figorbita01}
\end{figure}

In this section of this paper, we worked the MPD Equation not so much with
the Christoffel symbols, but with the RRC, which describe the trajectories
of spinning test particles around on rotating massive bodies. Now, we shall
compare numerically the trajectories of the test particles both the spinless
and spinning. For this comparison, first we need to compare between the
trajectories of the spinless test particles given by the Carter\'{}s equations with the trajectories given by MPD equations. In this case, we
take as parameters: radius constant $r=10,$ $E=0.9525$, $a=1$, $Q=4.224806$, 
$L=2.810974$, maximum latitude = 53.92928$
{{}^\circ}$ and magnitude of spin equal to zero. We replace these values in the
Equations of Carter (\ref{1}) - (\ref{4}) and find the initial values of the
four velocity: $dt_{0}/d\tau =1.00748$, $d\theta _{0}/d\tau =2.854\times
10^{-3}$, and $d\varphi _{0}/d\tau =8.255\times 10^{-1}$. Then, these
initial values of the four velocity are replaced in the set of MPD equations
for spinless test particles (\ref{4.11a}), (\ref{4.11c}) and (\ref{4.11d}).
We plot in 3D the trajectory for a spinless test particle in Mathematica
(Figure \ref{figorbita01}). Finally, we compare the data table of the cartesian
coordinates both the Carter\'{}s equations and MPD equations, and find that the difference of values is in the order of $10^{-7}$. 

\begin{figure}[h]
\centering
\includegraphics[width=0.6\textwidth]{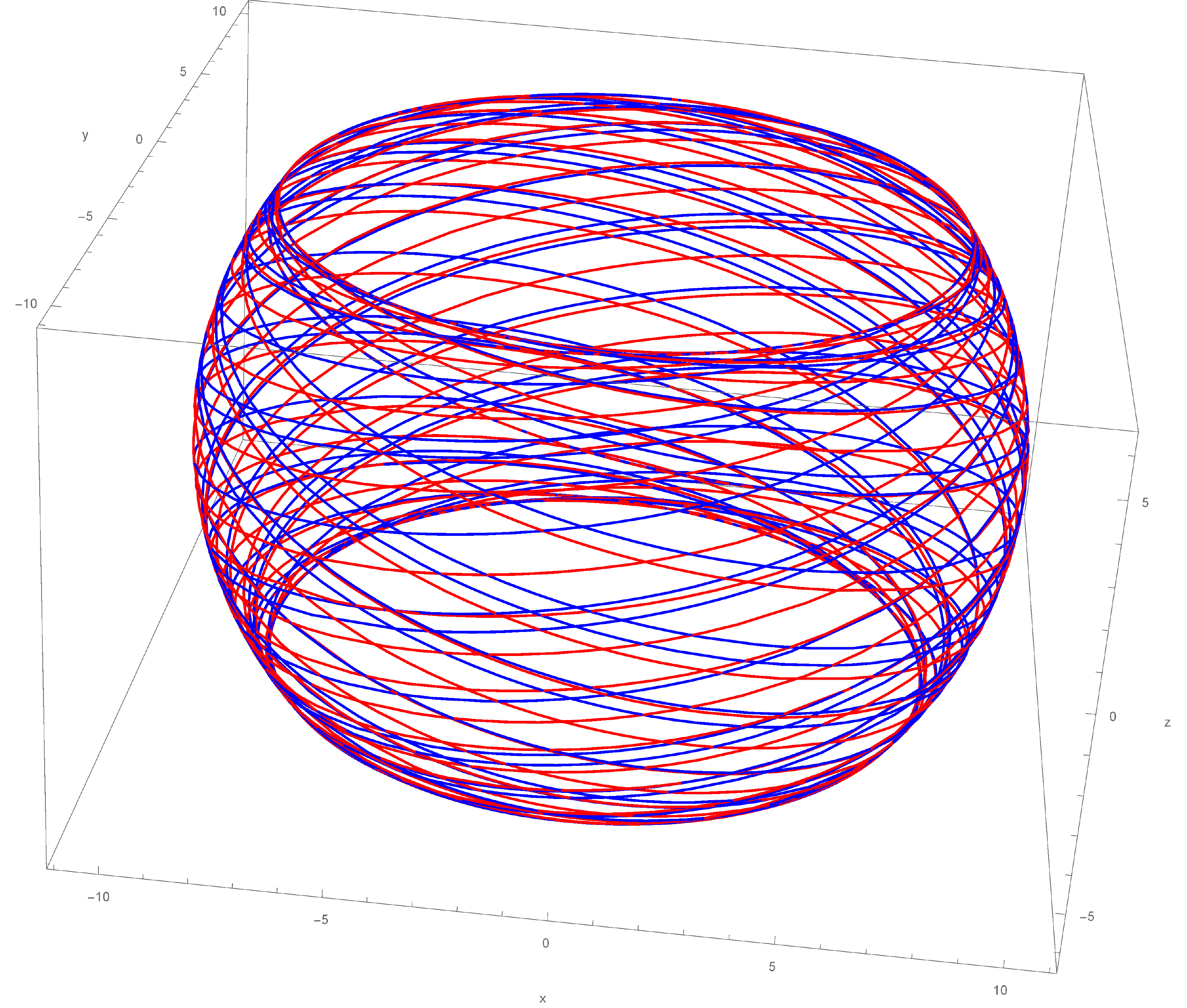}
\caption{Trajectories for spinless
(blue color) and spinning test particles (red color). The values of the
parameters are: $r=10$, $M=1$, $E=0.9525$, $a=1$, $Q=4.224806$, \thinspace $
L=2.810974.$}\label{fig2}
\end{figure}

Next step, we compare between the trajectory of a spinless test particle (
\ref{4.11a}, \ref{4.11c}, \ref{4.11d}), with the trajectory of a spinning
test particle (\ref{s1}, \ref{s2}) in non-equatorial planes via MPD
equations. We replace the values of the RRC and the components nonvanishing
of curvature tensor given by Mino \cite{mino}. Then, we plot in 3D the
trajectories both for a spinless test particle and a spinning test particle
in Mathematica (Figure \ref{fig2}). And finally, we compare the data table of
the cartesian coordinates both for a spinless test particle and a spinning
test particle, and find that the difference of values is in the order of $
10^{-4}$. This small difference in the trajectories is given by the effect
of spin in a test particle \cite{mash}.

\begin{figure}[h]
\centering
\includegraphics[width=0.6\textwidth]{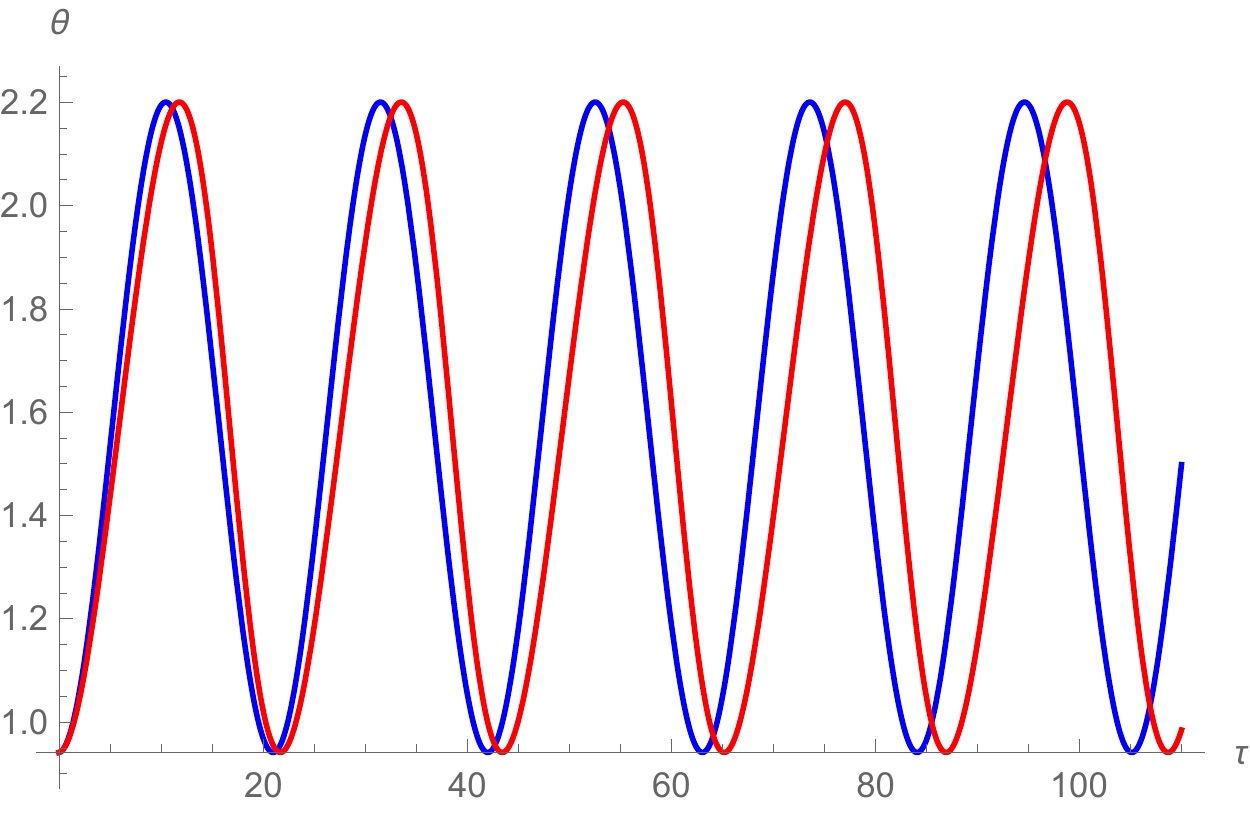}
\caption{Graph of polar angle ($\protect
\theta $) vs proper time ($\protect\tau $) for spinless (blue color) and
spinning test particles (red color). The initial conditions are: $\protect%
\theta _{0}=0.9412$, $d\protect\theta _{0}/d\protect\tau =0.03253$, \thinspace $
L=2.810974.$}\label{fig3}
\end{figure}

\begin{figure}[h]
\centering
\includegraphics[width=0.6\textwidth]{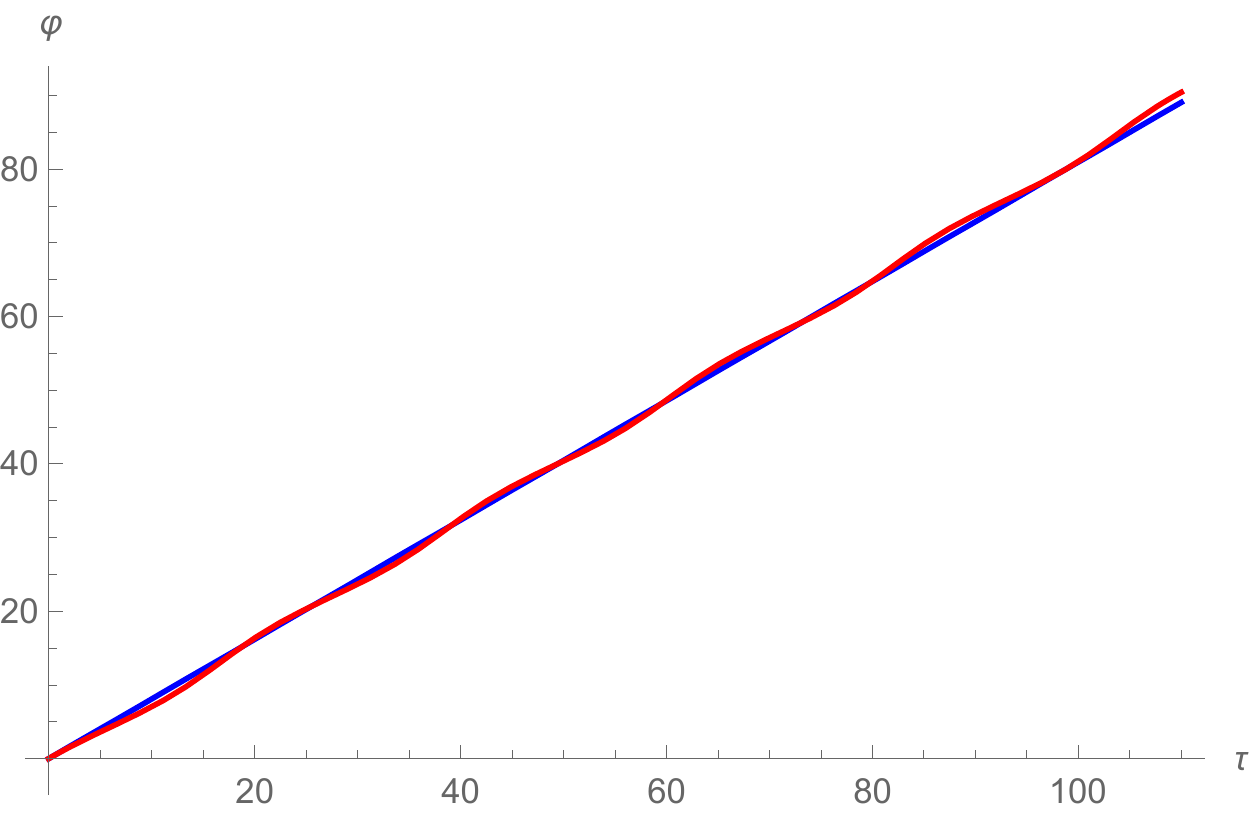}
\caption{Graph of azimuthal angle ($
\protect\varphi $) vs proper time ($\protect\tau $) for spinless (blue
color) and spinning test particles (red color). The initial conditions are: $
\protect\varphi _{0}=0$, $d\protect\varphi _{0}/d\protect\tau =0.08255$, \thinspace $
L=2.810974.$}\label{fig4}
\end{figure}

We make another numerical comparison of our results given by the MPD
equations in regard to the trajectories of spinless and spinning test
particles. In this case, we take the same initial values for the four
velocity given above which were obtained by the Carter\'{}s equations (\ref{1}) - (\ref{4}), and replace these initial values in the
MPD equations (\ref{s1}, \ref{s2}). Then, we take two Boyer Lindquist
coordinates ($\theta $, $\varphi $) and obtain a graph for each coordinate
in regard to the proper time ($\tau $) which show the difference in the
trajectories between the spinless and spinning test particles. The first one
is the figure of the polar angle versus proper time ($\theta $ vs $\tau $)
(Figure \ref{fig3}). In this graph, we find the period of the spinning test
particle is longer than the period of the spinless test particle. Moreover,
in each orbit, the difference between the trajectories of each test particle
increases, that is, it is an accumulative process. This is due to the
contribution of the value of spin in the trajectory. The test particles
starts in the polar initial angle ($\theta _{0}=53.92928{{}^\circ}$) and reaches the maximum value in the opposite side of the equatorial
plane ($\theta _{op}=143.2928{{}^\circ}$). One can see in this graph that both the spinless and spinning test
particle reaches the same maximum o minimum of latitude, but with a time
difference for the two paths. This is due to the precession of the spinning
test particle. The second figure shows the relation between the azimuthal
angle and the proper time (Figure \ref{fig4}). We find the slope of the graph
for the spinning test particle is winding with relation to the graph of the
spinless particle. This phenomenum is another effect of the spin of the
particle \cite{mash-1984} and like the previous graph this winding is due to
the nutation of the spinning particle. The MPD equations described with the
help of the RRC allow characterizing these phenomena.

\section{Conclusions}

We found that the equations of motion for spinning test particles in the MPD
equations (\ref{4.8a} - \ref{4.8d}) describe the relation between the spin
of the particle and the angular momentum of the central mass. For this case,
we take the variation of spin vector in the time and the first term includes
as much from the density of angular momentum ($a$) as from the spin of the
test particle ($S$). For the case of spin - orbit perturbation, the
equations yields the relation between the spin of the particle given by the
magnitude of spin ($S$) and the orbit given to both the curvature tensor ($
R^{\alpha }{}_{\beta \gamma \lambda }$) and the velocity of center of mass ($
v^{\mu }$).

For circular orbits close to equatorial plane, the difference of
trajectories between spinless test particles and spininng test particles is
very small; but when the particles have spin the difference in the time is
bigger that when the particles does not have it.

The majority of works that include the Christoffel symbols in the MPD
equations does a description of the coupling between spin and field while in
this paper, since that we include the RRC in the MPD equations, we study not
only this coupling, but also the rotation of the frame that travels with the
test particle. These phenomena are studied with graphs both of the polar
angle and the azimuth angle.

Additionally the results obtained in this paper may be important guideline
to quantitative estimates of gravitational waves in numerical relativistic
simulations. When there is a spinning test particle in a gravitational field
the description of its geodesics is given, in the same time, the
characteristics of the space-time for where is traveling the particle. Now,
if the gravitational waves are detected for instruments as the Laser
Interferometric Gravitational Wave Observatory (LIGO) or the Laser
Interferometer Space Antenna (LISA), it is possible comparating the
description of the orbits of the spinning test particles with the signal of
these instruments and giving some conclusions. In the majority of these
researches, they use techniques in interferometry. Our results are good
material when it comes to face the construction of a gravitational wave
detector that takes account the effect of spin in the template of
interferometry. The description of this phenomenom can be an extension of
the work done in this paper with help of the MPD equations.

\bmhead{Acknowledgments}
The authors are grateful with Pontificia Universidad Javeriana at Bogot\'{a}.
\begin{appendices}

\section{The Ricci rotation coefficients of the Kerr spacetime}\label{secA1}




The nonvanishing components are

\begin{equation}
\omega _{01}{}^{0}=\omega _{00}{}^{0}=\omega _{1}\text{, \ \ }\omega _{1}:=
\frac{1}{\Sigma ^{{{}^3}/{{}^2}}\Delta ^{{\frac12}
}}\left( ra^{2}\sin ^{2}\theta -Mr^{2}+Ma^{2}\cos ^{2}\theta \right) , 
\tag{A 1}
\end{equation}

\begin{equation}
\omega _{31}{}^{0}=\omega _{30}{}^{1}=\omega _{13}{}^{0}=\omega
_{10}{}^{3}=\omega _{03}{}^{1}=-\omega _{01}{}^{3}=\omega _{2}\text{, \ \ }
\omega _{2}:=\frac{ar\sin \theta }{\Sigma ^{{{}^3}/{{}^2}
}},  \tag{A 2}
\end{equation}

\begin{equation}
\omega _{22}{}^{1}=-\omega _{21}{}^{2}=\omega _{33}{}^{1}=-\omega
_{31}{}^{3}=\omega _{3}\text{, \ \ }\omega _{3}:=\frac{r\Delta ^{
{\frac12}
}}{\Sigma ^{
{{}^3}/{{}^2}}},  \tag{A 3}
\end{equation}

\begin{equation}
\omega _{02}{}^{0}=\omega _{00}{}^{2}=\omega _{12}{}^{1}=-\omega
_{11}{}^{2}=\omega _{4}\text{, \ \ }\omega _{4}:=\frac{a^{2}\cos \theta \sin
\theta }{\Sigma ^{{{}^3}/{{}^2}}}  \tag{A 4}
\end{equation}

\begin{equation}
\omega _{32}{}^{0}=\omega _{30}{}^{2}=-\omega _{23}{}^{0}=-\omega
_{20}{}^{3}=\omega _{03}{}^{2}=-\omega _{02}{}^{3}=\omega _{5}\text{, \ \ }
\omega _{5}:=\frac{a\cos \theta \Delta ^{{\frac12}
}}{\Sigma ^{{{}^3}/{{}^2}}}  \tag{A 5}
\end{equation}

\begin{equation}
\omega _{33}{}^{2}=-\omega _{32}{}^{3}=\omega _{6},\ \ \omega _{6}:=\frac{
\left( r^{2}+a^{2}\right) \cos \theta }{\Sigma ^{{{}^3}/{{}^2}}\sin \theta }  \tag{A 6}
\end{equation}
\\
Here it is followed the notation $\omega _{1}-\omega _{6}$ \cite{tanaka-1996}
\ and grouped the components of same value with a simple index $\omega _{i}$ 
$\left( i=1-6\right) $. Where: $a$ is the angular momentum of the source,$\
\Sigma :=r^{2}+a^{2}\cos ^{2}\theta $ and $\Delta :=r^{2}+a^{2}-2Mr$.

\end{appendices}


\bibliography{sn-bibliography}


\end{document}